\documentclass[twocolumn,notitlepage,aps,prb,10pt,superscriptaddress,floatfix,letterpaper,reprint
]{revtex4-2}	

\usepackage{hyperref}
\usepackage[usenames,dvipsnames]{color}
\usepackage{amsmath}
\usepackage{mathtools}
\usepackage[english]{babel}
\usepackage{amssymb}
\usepackage{graphicx}
\usepackage{epstopdf}
\usepackage[normalem]{ulem}
\usepackage{subfigure}
\usepackage{todonotes}
\usepackage{braket}
\usepackage{bbold}
\usepackage{placeins}
\usepackage[para]{threeparttable}
\usepackage{lipsum}
\usepackage{multirow}
\usepackage{bm}

\definecolor{linkcolor}{HTML}{399B03}
\definecolor{urlcolor}{HTML}{399B03}
\hypersetup{pdfstartview=FitH, linkcolor=linkcolor,urlcolor=urlcolor,colorlinks=true}
\hypersetup{
    unicode=false,          
    pdftoolbar=true,        
    pdfmenubar=true,        
    pdffitwindow=false,     
    pdfstartview={FitH},    
    pdfauthor={},         
    colorlinks=true,        
    linkcolor=blue,         
    citecolor=red,          
}

\begin{document}

\title{Real-time propagation of adaptive sampling selected configuration interaction wave function}
\author{Avijit Shee}
\affiliation{Department of Chemistry, University of California, Berkeley, USA}
\author{Zhen Huang}
\affiliation{Department of Mathematics, University of California, Berkeley, CA 94720, USA}
\author{Martin Head-Gordon}
\affiliation{Department of Chemistry, University of California, Berkeley, USA}
\author{K. Birgitta Whaley}
\affiliation{Department of Chemistry, University of California, Berkeley, USA}

\begin{abstract}
We have developed a new time propagation method, time-dependent adaptive sampling configuration interaction (TD-ASCI), to describe the dynamics of a strongly correlated system. 
We employ the short iterative Lanczos (SIL) method as the time-integrator, which provides a unitary, norm-conserving, and stable long-time propagation scheme. 
We used the TD-ASCI method to evaluate the time-domain correlation functions of molecular systems. The accuracy of the correlation function was assessed by Fourier transforming (FT) into the frequency domain to compute the dipole-allowed absorption spectra. The FT has been carried out with a short-time signal of the correlation function to reduce the computation time, 
using an efficient alternative FT scheme based on the ESPRIT signal processing algorithm. We have applied the {TD-ASCI} method to prototypical strongly correlated molecular systems and compared the absorption spectra to spectra evaluated using the equation of motion coupled cluster (EOMCC) method with a truncation at single-doubles-triples (SDT) level.      
\end{abstract}


\maketitle

\section{Introduction}

An explicit solution of the time-dependent Schr{\"o}dinger equation opens up numerous new possibilities for studying electron dynamics in many-body systems. This includes ultrafast charge and energy migration \cite{CEDERBAUM1999205, Kuleff_2014, KuleffPRL07}, as well as various spectroscopic techniques, such as photoionization \cite{photoionize_chuPhysRevA04, photoionize_deWijn_2008}, X-ray absorption \cite{xrayKadekPCCP15, xrayLi18, xrayLopataJCTC12}, and valence electron UV/vis spectroscopy \cite{optical_LiJCP07, UVVis_LiPCCP05}. In spectroscopic methods, electronic responses are typically studied in the presence of a time-dependent external electromagnetic field, from which molecular information about energy eigenstates is extracted. While frequency-domain approaches, commonly used in the quantum chemistry community, can also extract 
such information, these face several challenges: the computational method scales linearly with the number of requested roots, and finding interior roots is often difficult \cite{zuevJCC15, PengJCTC15}. In contrast, time-domain methods allow us to extract spectral information over a broad frequency range from the resulting signals. 
We focus here on evaluating one such signal, time-dependent correlation functions, specifically the dipole-dipole autocorrelation function, from which a system's linear absorption spectrum can be obtained.

Just as for time-independent electronic structure theory, in time-domain methods, the underlying description of electron correlation
plays a crucial role. The choice of which method is used for this determines: a) whether the time evolution operator will be unitary so that the wave function norm is conserved during long-time evolution, and b) whether the strongly correlated nature of the ground state and the propagated wave function is adequately captured. Amongst the high-accuracy electronic structure methods, 
configuration interaction (CI) \cite{TDCI_Greenman_PRA10, TDCI_Wilson_JCP2018} and 
coupled cluster (CC) \cite{sverdrup2023time,pedersen2019symplectic} wave functions have been 
used for time propagation in 
recent works. However, when low excitation rank truncation is employed, both CI-based and CC-based time propagation fail in the strongly correlated regime. Additionally, apart from the simple CIS method, CI-based time propagation is unreliable as the system size is increased, while CC-based time propagation fails to maintain unitarity, particularly when the wave function becomes strongly correlated. Another time propagation method, the time-dependent density matrix renormalization group (TD-DMRG) \cite{TD_DMRGBaiardi21}, has achieved greater success than the previously mentioned methods. Nevertheless, because the area law of entanglement, which underlies the efficiency of DMRG ground state calculations,
often does not hold for time-propagated wave functions, the success of time-dependent DMRG is somewhat limited compared to time-independent frameworks \cite{TDDMRGPRA20, TDDMRGPRL20}. 
These experiences motivate the development of an accurate and efficient time propagation method for a strongly correlated wave function.
In this work, we explore the time propagation of another wave function, namely the adaptive sampling configuration interaction (ASCI) wavefunction \cite{tubman2020modern, williams2023parallel}, which can accurately describe strongly correlated ground state wave functions. We refer to this method as TD-ASCI. We shall develop the TD-ASCI methodology that provides unitary electronic dynamics and will explore its ability to describe strong correlation of molecular systems during time propagation.
In this first work we have selected several prototypical strongly correlated molecular systems for detailed study, to benchmark the method and analyze its numerical features. In the future, we shall also address extensions and broader applications of the method to reduced dynamics for large numbers of electrons and to open quantum systems dynamics in which electronic excitations are coupled to non-Markovian vibrational degrees of freedom.

The choice of numerical integration technique for time propagation is of vital importance for this work. The widely used fourth-order Runge-Kutta (RK4) does not obey the symplectic nature of the time-propagation. Various other schemes, in particular, the split operator (SO) technique \cite{SO_FEIT1982412} and the second order difference (SOD) method \cite{KOSLOFF198335} employed with TD-CC and TD-CI methods also suffer from the same issue, especially for long-time propagation. Therefore in this work, we shall employ the short iterative Lanczos (SIL) time-integration scheme first proposed by Park and Light \cite{ParkLight_JCP86}. The SIL procedure provides a unitary, symplectic time-integration scheme which can be applied for long-time dynamics with a suitable update of the Krylov subspace. A related time-integration scheme is the Chebyshev orthogonal polynomial-based integrator (CH) \cite{ChebyshevJCP1984accurate}, which provides a compact global propagator. The accuracy and computational effort of the CH scheme is comparable with that of the method. But, in contrast, the CH scheme is not unitary, is valid only for Hermitian Hamiltonians, and is limited to time-independent Hamiltonians.         

Despite the advantages of the time-domain methods mentioned above, application of all of these to obtaining high-resolution absorption spectra is limited by the fact that the simulated signal requires a long-time evolution. Many efforts have been made to obtain high-resolution spectra from short-time signals. These include methods such as the filter diagonalization technique \cite{FD_NeuhauserJCP95}, the Pad{\'e}-based \cite{padeFT_BrunerJCTC16} FT, and the machine learning(ML)-based \cite{MLFT_HaugeRepiskyJCTC23} FT. In this work, we proposed a new technique to obtain spectra from short-time signals that is inspired by the signal processing technique ESPRIT \cite{roy1989esprit}, which has been found to be successful in reducing the circuit depth of quantum phase estimation (QPE) algorithms \cite{Stroeks_2022}.

The rest of the manuscript is organized as follows. In Section~\ref{sec:asci_gs} we first summarize the ASCI algorithm to prepare the ground state.  In Section~\ref{sec:lanczos} we outline our implementation of the Lanczos-based time evolution algorithm, 
with a particular emphasis on how the long-time dynamics is carried out.
In Section~\ref{sec:dipole_propagation} we present the algorithmic details required for evaluation of the dipole autocorrelation function.
In Section~\ref{Sec:FT} we present the novel FT scheme of the short-time signal based on ESPRIT.  Section~\ref{sec:application} then presents simulations of the absorption spectra of several multireference molecular systems that demonstrate the efficiency of the dynamics scheme and of the novel FT scheme. Section~\ref{sec:conclusion} summarizes and concludes with an outlook for future applications.

\section{Adaptive Sampling Configuration Interaction (ASCI)} \label{sec:asci_gs}
ASCI approximates the FCI expansion of a wave function: $| \Psi \rangle = \sum_I^N C_I \Phi_I$, by a dominant set of Slater determinants ($\Phi$s): $\{D\}$. The selection of those dominant determinants is done in an iterative procedure, where we try to generate a new determinant space from the old one by finding the connection through the Hamiltonian, $H$. At a given iteration step, $k$, the ASCI procedure 
selects new determinants not contained in the current step by using the following equation:

\begin{equation}
    C_I = \sum_{J \neq I} \frac{H_{IJ}C_J}{H_{II}-E_k} \label{Eq:ranking}
\end{equation}

Eq. \eqref{Eq:ranking} shows that starting from a trial wave function $|\Psi_t^k\rangle = \sum_J C_J \Phi_J $, we generate a new set of determinantal coefficients $C_I$s by using the first-order wave function correction formula. Here, $H_{IJ} = \langle \Phi_I | H | \Phi_J \rangle$, and $E_k$ is the energy of the trial wavefunction. The Hamiltonian, $H$, has a maximum excitation rank of 2, thus, the new determinant set ${\Phi_I}$ generated from $|\Psi_t^k\rangle$ are only singles and doubles (SD) excited determinants w.r.t. the existing set of ${\Phi_J}$. Therefore, we will not span the entire Hilbert space by the newly generated ${\Phi_J}$s. Moreover, the $H_{IJ}$ matrix is significantly sparse, implying that not all $C_I$s generated at this iterative step are significant. We rank the new determinant coefficients $C_I$s to choose only the significant ones (within some tolerance) until we reach a fixed target size of the determinant space, $N_{tdet}$. The ASCI algorithm also utilizes another parameter $N_{cdet}$, which limits the number of core determinants from the set ${|C_J\rangle}$, from which the new SD excited determinants are formed. This second parameter makes the search procedure more tractable.

The Hamiltonian is then projected onto the determinant space generated at each iterative step and is then diagonalized to obtain the total energy, $E_k$. We terminate the iteration procedure when we reach a desired level of convergence in terms of the total energy, $E_k$.

\section{Time dependent CI algorithm} \label{sec:lanczos}

Configuration Interaction (CI) is a wave function-based many-body method, in which the exact correlated wave function $|\Psi \rangle$ is expressed in terms of a linear expansion of the ground and excited Slater determinants $|\Phi_I\rangle$s: $|\Psi \rangle = \sum_I^N C_I |\Phi_I \rangle$, where $I$ stands for various electronic configurations. When $|\Psi \rangle$ is inserted
into the time-dependent Schr{\"o}dinger equation, we obtain the following equations of motion for the $C_I$ coefficients:  

\begin{equation}
    i \frac {\partial C_I}{\partial t} = \sum_{J=1}^N H_{IJ}(t) C_J \label{eq:TDCI}
\end{equation}
The solution of Eq.~\eqref{eq:TDCI} can be written as:
\begin{align}
    C_I (t_n) ={}& \sum_{J=1}^N U_{IJ} (t_n, t_0) C_J(t_0) \\
    U_{IJ} (t_n, t_0) ={}& \mathcal{T}\{\exp[-i\int_{t_0}^{t_n} dt H_{IJ}(t)]\}, \label{Eq: TDCIsoln}    
\end{align}

where $\mathcal{T}\{\dots\}$ is the time-ordering operator.

To evaluate $U_{IJ} (t_n, t_0)$ numerically we require two components: i) implementation of the time-ordering, and ii) the evaluation of the exponential of a large matrix. Since in this work, we restrict our implementation to the time-independent Hamiltonian, the time argument will henceforth be dropped from the Hamiltonian and implementation of the time ordering is not required. 

To avoid the cost of the matrix exponentiation of a large matrix, we first make a polynomial approximation of the time-evolution operator \cite{ParkLight_JCP86} and truncate this at order $p$:

\begin{align}
    C_I (t+\tau) ={}& \sum_{J=1}^N e^{-iH_{IJ} \tau} C_J(t) \nonumber \\
                 \approx & \sum_{J=1}^N \sum_{k=0}^p \frac {(-i\tau)^k}{k!} H_{IJ}^k C_J(t).
\end{align}
We then identify the truncation $p$ as the dimension $N_{Krylov}$ of a Krylov subspace spanned by vectors $\mathbf{a}^k$ defined by the following recursion relation:
\begin{equation} \label{eq:ak}
    \mathbf{a}^k = \mathbf{H} \mathbf{a}^{k-1} = \mathbf{H}^k \mathbf{a}^{0} ; \quad \mathbf{a}^0 = \mathbf{C}(t),
\end{equation}
resulting in an expansion of the time-dependent coefficients $ C_I (t+\tau)$ on the Krylov subspace:
\begin{align}
 C_I (t+\tau)    
                 \approx {}& \sum_{k=0}^{N_{Krylov}}\frac {(-i\tau)^k}{k!} a_I^k(t).
                 \label{Eq:polynomial}
\end{align}

The time propagation begins with $\mathbf{C}(0) = {\mathbf{a}}^0$ set equal to coefficients of the CI ground state $|\Psi\rangle_{CI}$, which is of dimension $N$.
The $\mathbf{a}^k$ vectors span a Krylov subspace $V_p=\text{span}(\{\mathbf{a}^0, \mathbf{a}^1, \hdots \mathbf{a}^{p-1}\})$. Let us define $A_p = [\mathbf{a}^0 \mathbf{a}^1 \hdots \mathbf{a}^{p-1}]$. If we employ the Lanczos procedure to generate the subspace (see Appendix~\ref{sec:lanczos_detail}), the Hamiltonian matrix $\mathbf{H}$ then becomes tri-diagonal within that subspace:
\begin{equation}
    \mathbf{H}_p = Q_p^\dag \mathbf{H} Q_p=   \begin{pmatrix}
      \alpha_1 & \beta_1 & 0 & ... & 0 \\
      \beta_1 & \alpha_2 & \beta_2 & ... & 0 \\
      0       &  \beta_2 & \alpha_3 & \ddots &   \\
      \vdots  &  \vdots      & \ddots  & \ddots & \beta_{p-1} \\
       0      &  0 &            & \beta_{p-1} & \alpha_p
   \end{pmatrix} \label{Eq:tridiag},
\end{equation}
where $\alpha_k$ and $\beta_k$ ($k=1,\cdots,p$) are as defined in Appendix~\ref{sec:lanczos_detail} and $Q_p$ is obtained from the reduced QR factorization of $A_p$: $A_p = Q_p R_p$. The matrix $Q_p$ is composed of the vectors $\mathbf{q}^k$ ($k=1,\cdots,p$, see Appendix~\ref{sec:lanczos_detail} for the definition): $Q_p = [\mathbf{q}^1 \mathbf{q}^1 \hdots \mathbf{q}^{p}]$.

Each vector added to the subspace $V_k$ is orthogonalized against all previous vectors present in the subspace.
We use a Gram-Schmidt procedure for this orthogonalization step because of its modest $\mathcal{O} (N_{Krylov}^2 N)$ scaling, where $N_{Krylov}$ is the size of the Krylov space, and $N$ is the number of determinants in the CI expansion. This step may become as expensive as the $\sigma$-vector (matrix-vector product) evaluation step in Eq.~ \eqref{eq:ak}
when the size of the Krylov space is large, since the latter scales as $\mathcal{O} (N_{Krylov} N^2)$.


We can use the matrix $Q_p$ to make a basis transformation of the coefficients $C_J(t)$, i.e., $\mathbf{d}(t) = Q_p^\dag \mathbf{C}(t)$, where $\mathbf{d}(t)$ is a ($p\times 1$) vector. 
Substituting in Eq.~ \eqref{eq:TDCI} and using the tridiagonal form of the Hamiltonian in Eq.~ \eqref{Eq:tridiag} generates 
a reduced dimensional ($p\times p$) time-evolution equation.
The solution of the reduced dimensional equations can then be written down as

\begin{equation}
     \mathbf{d}(t+\tau) = e^{-i\mathbf{H}_p \tau} \mathbf{d}(t), \label{Eq:evolve_Lanczos}
 \end{equation}

Since $\mathbf{H}_p$ is tri-diagonal and of very small dimension we can easily diagonalize $\mathbf{H}_p$ and can evaluate the exponential factor in Eq. \eqref{Eq:evolve_Lanczos} as $T e^{-i \mathbf{\Lambda}_p \tau } T^\dag$, where $T$ is the transformation matrix that diagonalizes $\mathbf{H}_p$ and $\mathbf{\Lambda}_p$ is the diagonal matrix consisting of the eigenvalues from that diagonalization. The CI coefficient vector $\mathbf{C}$(t+$\tau$) is then obtained as:

\begin{equation}
    \mathbf{C}(t+\tau) = Q_p \mathbf{d}(t+\tau).\label{Eq:krylovtoCI}
\end{equation}

\subsection{Refreshing the subspace for long time propagation:} \label{sec:refresh}

The propagator constructed above with the subspace Hamiltonian $\mathbf{H}_p$ does not remain valid for long times. 
The Taylor expansion of Eq. \eqref{Eq:krylovtoCI} is given by:

\begin{equation}
   \mathbf{d}(t+\tau) =\sum_{k=0}^{\infty} \frac {(-i\tau)^k}{k!} (\mathbf{H}_p)^k \mathbf{d} (t),\label{Eq:small_conv}
\end{equation}
where the first $p$ terms of this series coincides with that of the Taylor expansion of $Q_p^\dagger \mathrm e^{-\mathrm i \mathbf H\tau}\mathbf C(t)$ (see Eq. \eqref{Eq: TDCIsoln}) due to the Krylov construction.
Thus we control the error by ensuring that the norm of the $p$-th order term $\mathbf{d}^{(p)}(t,\tau)$ of Eq. \eqref{Eq:small_conv} is bounded by $\epsilon$, i.e.,
\begin{equation}
    \|\mathbf{d}^{(p)}(t,\tau)\| \leq \epsilon \label{Eq:conv_threshold},
\end{equation}    
where
\begin{equation}
    \mathbf{d}^{(p)}(t,\tau) = \frac {(-i\tau)^p}{p!} (\mathbf{H}_p)^p \mathbf{d} (t).
\end{equation}

Since $\mathbf d(t)$ is a unit vector, from Eq. \eqref{Eq:conv_threshold}, then 
$\|\mathbf{d}^{(p)}(t,\tau) \|\leq \frac{\tau^p}{p!}\|\mathbf H_p\|_2^p$, where $\|\mathbf H_p\|_2$ is the matrix 2-norm of $\mathbf H_p$, i.e., the maximal eigenvalue of $\mathbf H_p$ since $H_p$ is a positive Hermitian matrix. Therefore,
the maximum possible time interval $\tau_{max}$ is given by


\begin{equation}
    \tau_{max} = \left(\epsilon \left[\frac{p!}{\|\mathbf H_p\|_2} \right]\right)^\frac{1}{ p}, 
\end{equation}
This means that for a given time $t$, the value of the next time interval, $\tau_{max}$, depends on the current value of the quantum state expressed in the Krylov space by the vector $\mathbf{d}(t)$ and the tridiagonal matrix $\mathbf{H}_p$. Therefore we have an adaptive time grid. This aspect will be further explored in Section \ref{sec:dipole_propagation}.

This analysis 
implies that the time-evolution operator $U_p = e^{-i\mathbf{H}_pt}$ defines a global propagator over the time interval $\left[t, t+\tau_{max} \right]$. 
After the time ($t+ \tau_{max}$), the time-evolution operator $U_p$ becomes less accurate and we need to construct a new $\mathbf{H}_p$. To that end, we first build a new vector of CI coefficients at 
time ($t+\tau_{max}$), i.e., $\mathbf{C}$(t+$\tau_{max}$), from Eq. \eqref{Eq:krylovtoCI}. 
This vector is then used as an initial vector to construct a new Krylov space, $V_p$, and a new tridiagonal Hamiltonian, $\mathbf{H}_p$, in that space, in a manner analogous to Eqs.~\eqref{eq:ak} and \eqref{Eq:tridiag}.
To achieve a long time propagation, this procedure is repeated for many time intervals $\left[t, t+\tau_{max} \right]$, which provides a very stable propagation scheme. 




\section{Evaluation of the dipole auto-correlation function} \label{sec:dipole_propagation}
The correlation function between two operators $A$ and $B$ is defined as: 
\begin{equation}
  C_{AB} (1, 2) = \langle A_H(1) B_H(2) \rangle \label{Eq:corr_function} ; \quad 1, 2 \equiv (x_1, t_1), (x_2, t_2),
\end{equation}
where $A_H(1)$ and $B_H(2)$ are operators expressed in the Heisenberg representation, i.e.,
\begin{align}
   A_H(1) = {}& e^{-iHt_1} A (x_1) e^{iHt_1} \\
   B_H(2) = {}& e^{-iHt_2} B (x_2) e^{iHt_2} .
\end{align}
From the correlation function, we can obtain the experimentally relevant linear response function:

\begin{align}
    \chi_{AB} (1,2) = {}& -i \Theta (t_1 - t_2) \langle [A_H(1), B_H(2)] \rangle \nonumber \\
                    = {}& -i \Theta (t_1 - t_2) ( C_{AB}(1,2) - \Tilde{C}_{AB} (1,2)),
\end{align}
where $\Tilde{C}_{AB}(1,2)= \langle  B_H(2) A_H(1) \rangle$ is the commuted correlation function as defined in Eq. \eqref{Eq:corr_function}. 
$\chi_{AB} (1,2)$ measures the change in expectation value of an operator $A_H(1)$ to linear order in the presence of an external field $\phi(2)$, where the external field interacts with an operator $B(x_2)$ on the system via a perturbation $\int dx_2 \phi(2) B(x_2)$. The evaluation of $\chi_{AB} (1,2)$ allows us to avoid adding an external perturbation explicitly to the Hamiltonian, thus providing a more stable time-propagation scheme, 
while also allowing extraction of experimentally observed quantities. In this work, we 
restrict ourselves to the perturbation where the electric dipole of a system interacts with an external electric field $\hat{E}(x,t)$ following the dipole approximation: $V(x, t) = -\hat{\mu} (x)\cdot\hat{E}(x,t)$. 
To measure the first-order change in dipole moment we then have to evaluate the following correlation function:

\begin{align}
C_{\mu \mu}(t) ={}& \langle \mu_H(t) \mu_H(0)\rangle \nonumber \\
        ={}& \langle \Psi | e^{i H t} \hat \mu e^{-iHt} \hat \mu |\Psi\rangle \label{Eq:dipole_auto}, 
\end{align}
where we have integrated out the spatial degrees of freedom. From now on we will carry only the time index for the correlation function. 

In Eq. \eqref{Eq:dipole_auto}, we will use the second quantized definition of the dipole operator to be consistent with an implicit second quantized representation of $|\Psi\rangle$ used in the ASCI formulation. The key component for evaluating $C_{\mu \mu}(t)$ is then the time-propagation of a new vector:

\begin{equation}
    \mathbf{m} = \hat \mu |\Psi\rangle. ; \quad \hat \mu = \sum_{pq}\mu_{pq}p^\dag q
\end{equation}

We can now concisely write

\begin{equation}
   C_{\mu \mu}(t) = e^{i E_{gr} t} \langle \mathbf{m}(0) | \mathbf{m}(t) \rangle,
\end{equation}
where $\mathbf{m}(t) = e^{-iHt} \mathbf{m}(0)$, $e^{i E_{gr} t} \langle \Psi | = \langle \Psi | e^{i H t} $ and $E_{gr}$ is the electronic energy of the ground state of the system as evaluated from an electronic structure method. 

\begin{figure}
    \centering
    \includegraphics[width=0.8\linewidth]{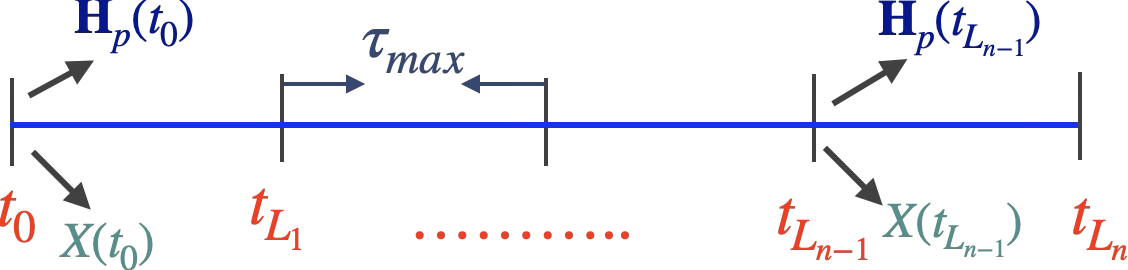}
    \caption{Schematic for the time propagation of the dipole vector, $\mathbf{m}$. The points $t_{L_i}$ represent the adaptive grid points obtained from the Lanczos algorithm, separated by variable time intervals $\tau_{max}$. We extract a new Lanczos Hamiltonian $\mathbf{H}_p(t_{L_i})$, and vector $X(t_{L_i}) = Q^\dag_p(t_{L_i}) \mathbf{m}(0)$ on each of the points $t_{L_i}$.  Exact interpolation of $C_{\mu \mu} (t)$ on any other time grid points not contained in $\{t_{L_i}\}$ can be readily performed using $\mathbf{H}_p$ and $X(t_{L_i})$, using minimal computation time. See text for more details.}
     
\label{fig:dipole_corr}
\end{figure}

We can further simplify the evaluation of $C_{\mu\mu}(t)$ by combining the time evolution scheme described in Section \ref{sec:lanczos} and the adaptive grid technique described 
in Section \ref{sec:refresh}.
We need to additionally evaluate only the tridiagonal matrix $\mathbf{H}_k$ of dimension $(N_{Krylov} \times N_{Krylov})$, 
as well as a vector $X = Q_k^\dag (t_{L_i}) \mathbf{m}(0)$ of dimension ($N_{Krylov} \times 1$), on each adaptive grid point  $t_{L_i}$. 
Here, $Q_k(t_{L_i})$ 
is the transformation matrix between the Krylov space and the actual configuration space on each $t_{L_i}$. 
The 
overall scheme is depicted in Fig. \ref{fig:dipole_corr}. 
Since we already have a global propagator for the interval $t_{L_i} \leq t < t_{L_{i+1}}$, namely $e^{-i\mathbf{H}_kt}$ (Section~\ref{sec:refresh}), 
this suffices to evaluate any time-evolved vector $\mathbf{d}(t)$ within that interval in the corresponding reduced dimensional space for that interval. 
The correlation function can then be evaluated as $C_{\mu \mu}(t) = e^{i E_{gr} t} \langle X | \mathbf{d}(t) \rangle$ for any point within that time interval. Note that this avoids any interpolation error. The computational cost of the procedure is then dominated by the linear dependence on the cost of regenerating the Krylov subspace at each $t_{L_i}$ on the adaptive grid, and the cost of the time-propagation within each interval $\tau_{max}$ is negligible by comparison. 

When the Lanczos time propagation algorithm is expressed in this manner, it becomes equivalent to the orthogonal polynomial-based Chebyshev (CH) integrator, which is also free of interpolation error.

 However, unlike the SIL integrator where a grid is defined dynamically as we propagate, for the CH integrator the time grid for global propagation is defined beforehand from the nodes of the polynomials. It has been shown in the context of the TD-EOMCC method~\cite{TDEOM_CYang_JCTC24} that the number of computationally most expensive $\sigma$-vector evaluations (Eq.~ \eqref{eq:ak}) for the SIL integrator is fewer than for the CH integrator. 
 


\section{Correlation function from ASCI wave function propagation}

The starting point for the Lanczos time evolution in Eq. \eqref{Eq:polynomial} with the TD-ASCI method is the coefficient vector $\mathbf{C}_{ASCI}$ of the ASCI ground state estimation, as described in Sec. \ref{sec:asci_gs}. It excludes a large part of the determinant space (DS) through a robust selection procedure. However, the neglected part of the DS may contribute significantly during the time evolution. Therefore, a robust time-evolution scheme requires the evaluation of an optimal DS beyond just those determinants required for the ground state. In this context, we will first 
find the space required for evaluating the dipole autocorrelation function. The dipole operator, $\hat\mu = \sum_{pq}\mu_{pq}p^\dag q$, is a one-particle operator that can generate all singly excited states with respect to the ground state, $|\Psi \rangle$. The determinantal coefficients of the singly excited space generated by the dipole operator can be expressed as follows:

\begin{equation}
    a_I^{(J)} = \mu_{IJ} C_J ; \quad A_I = \sum_{J=1}^N a_I^{(J)},
\end{equation}
where $\mu_{IJ}$ is the dipole matrix element in the determinant basis.
We exploit the sparsity of the dipole matrix by retaining $|a_I^{(J)}| > \epsilon$ to obtain $A_I$, where $\epsilon$ is a predefined threshold. We then rank the $A_I$ values to select all dominant determinants contributing to the single excitation space, the number of which is upper bounded by the $N_{tdet}$ parameter, as described in Sec.~ \ref{sec:asci_gs}.
We point out that the selection procedure used here does not exclude the ground state determinants as is done in Eq. \eqref{Eq:ranking}. However, depending on the polarization of the dipole vector, the excited space may not contain the ground state determinants if it has a different symmetry than the ground state, which is ensured in this case by exploiting the sparsity as mentioned above. Furthermore, this ranking procedure does not include any preconditioner and thus it does not bias the selection to any particular state. Similar to the ground state procedure, we choose a selected set of core determinants from which the excited space is generated. This number turns out to be significantly larger than $N_{cdets}$ for the ground state. We will show numerically in Section \ref{sec:application} that the choice of this number of selected determinants is crucial to obtaining a converged result. In addition, as in the ground state preparation procedure, the choice of a one-particle basis to expand the determinants is also crucial. We also show the numerical implications of this choice in Section~\ref{sec:application} below.

Having obtained a new determinant space by operating with the dipole operator, we rebuild the Hamiltonian, $H_{IJ}$ within that space, and carry out time-propagation of the $A_I$ vector for that Hamiltonian, noting that the determinant space remains fixed for the entire time-evolution process.  The other numerical details of the time-propagation exactly follow the scheme described in Sec. \ref{sec:dipole_propagation}. 

\section{Fourier transformation of the short time signal} \label{Sec:FT}

The Fourier transformation of $\chi_{\mu \mu} (t)$ takes the following form:

 \begin{equation}
        \chi_{\mu \mu}(\omega) = \sum_\lambda \frac{\mu_{0\lambda} \mu_{\lambda 0}}{\omega - \omega_\lambda +i\eta} - \frac{\mu_{0\lambda} \mu_{\lambda 0}}{\omega + \omega_\lambda +i\eta}, \\
 \end{equation}
where $\lambda$ indexes the optically active excited states, $\omega_\lambda = E_\lambda - E_0$, and $\eta$ is a positive infinitesimal value. 
It can be shown that the absorption spectra is obtained from the Fourier transformed response function  according to \cite{mukamel1995principles, martin2016interacting}:

\begin{equation}
    A(\omega) = -\frac{1}{\pi} \mathrm {Im} \chi_{\mu \mu}(\omega)
\end{equation}
Now, from the fluctuation-dissipation theorem \cite{FDTheorem_Kubo_1966} at zero-temperature we obtain that $ \mathrm{Im} \chi(\omega) = \frac{1}{2} C_{\mu \mu} (\omega)$. Therefore, to calculate the absorption spectrum, we need to perform the FT of $C_{\mu \mu} (t)$. Evaluating this FT using the fast Fourier transform (FFT) requires a long-time signal, significantly increasing the computation time, as noted in previous works \cite{padeFT_BrunerJCTC16}. Many efforts have been made to obtain converged absorption spectra from short-time signals \cite{FD_NeuhauserJCP95, MLFT_HaugeRepiskyJCTC23, padeFT_BrunerJCTC16}. Among these, the most notable is the Pad{\'e}-based FT proposed by Bruner \textit{et al.} \cite{padeFT_BrunerJCTC16}.
Recall that
the discrete FT of the correlation function can be expressed as a power series:
\begin{equation}
    C_{\mu \mu}(\omega) =\Delta t \sum_{k =1}^{N_t} C_{\mu \mu} (k\Delta t) (e^{(i\omega -\gamma) \Delta t})^k.
\end{equation}
Here $N_t$ is the number of sample points and  $\gamma$ is a damping factor.
Introducing $z$ and $c_k$:
$$
c_k = \Delta t C_{\mu \mu} (k\Delta t),\quad z = e^{(i\omega -\gamma) \Delta t},
$$
we have $C_{\mu\mu}(\omega) = \sum_{k =1}^{N_t} c_kz^k$.
The Pad\'e approach aims to approximate this function via the following rational form:
\begin{equation}
  \sum_{k = 1}^{N_t} c_k z^k  = \frac{P(z)}{Q(z)}= \frac {\sum_{k = 1}^M a_kz^k}{\sum_{k = 1}^M b_k z^k} \label{Eq:pade_corr}.
\end{equation}
Here $M$ is the maximal order of the polynomials $P$ and $Q$, and  is chosen to be $\frac{N_t  - 1}{2}$.
The unknown parameters in Eq. \eqref{Eq:pade_corr}, i.e., $a_k$ and $b_k$, can be obtained via solving a linear equation system. The value of $C_{\mu \mu}(\omega)$ for any given $\omega$ 
can then be easily evaluated. The poles of $C_{\mu \mu}(\omega)$ can be found by calculating zeros of $Q(z)$. However, in this approach, the number of poles, which is equal to the order of the polynomial $Q$, is $M=\frac{N_t-1}{2}$, and is often too high. Thus, in the resulting spectra, we see many spurious roots.

In this work, we aim to find the spectra based on the idea that the time-domain correlation function can be expressed in terms of complex exponential functions and complex weights:

\begin{equation}
    C_{\mu \mu}(t) \approx \sum_k C_{k}^\mu   e^{-i\omega_k t},\quad C_{ k}^\mu , \omega_k\in\mathbb C.
\end{equation}
The first step in this algorithm is to find $\omega_k$. To that end, we use the Estimation of Signal Parameters via the Rotational Invariant Techniques (ESPRIT) algorithm \cite{roy1989esprit}. This technique has been recently employed in the context of fitting correlation functions in open quantum system problems \cite{ParkHuangZhuetal2024, BCF_ESPRITJCP24} and in quantum computing \cite{Stroeks_2022}. The ESPRIT algorithm first constructs a  Hankel matrix from the time-series data $C_{\mu \mu}(t)$ on a uniform time-grid with $\Delta t$ time-step:

\begin{equation}
   \mathbf M = \begin{pmatrix}
    C_{\mu \mu}(t_0) & C_{\mu \mu}(t_1)  & \cdots & C_{\mu \mu}(t_{n_t-1}) \\
    C_{\mu \mu}(t_1) & C_{\mu \mu}(t_2) & \cdots & C_{\mu \mu}(t_{n_t}) \\
    \vdots & \vdots &  \ddots  & \vdots \\
    C_{\mu \mu}(t_{n_t-1}) & C_{\mu \mu}(t_{n_t})  & \cdots & C_{\mu \mu}(t_{2n_t-2}) \\   
\end{pmatrix}
\end{equation}
With the Hankel matrix, we perform a singular value decomposition (SVD), i.e., $   \mathbf M = U \Sigma V$. We then obtain the $\omega_k$ values in two steps:
\begin{enumerate}
    \item evaluate the eigenvalues $\lambda_1, \lambda_2, \cdots, \lambda_L$ of  matrix $W = U_1' U_2$, where $U_1$, $U_2$ are defined as:
       \begin{equation}
            U_1 = U[1:n_t-1, 1:L],\quad  U_2=U[2:n_t, 1:L],
       \end{equation}
     and $U_1'$ is the Moore-Penrose pseudoinverse of $U_1$. Here $L$ is the chosen number of modes, which should be less than the number of non-zero singular values of $\mathbf M$.

    \item  evaluate
      \begin{equation}
          z_k = \mathrm i\frac{\log(\lambda_k)}{\Delta t} ;\quad   k = 1, \cdots,L,
      \end{equation}
       where log is the principal value of the logarithm.       
\end{enumerate}
We will then employ linear least square fitting to find $C_{k}^\mu$ for a given set of $z_k$.
Choosing a predefined value for the number of modes $L$ is often challenging, since the resulting spectra may not converge if the value of $L$ is too small. To choose a reasonable value for $L$, we apply a low-pass filter to the time domain signal, $C_{\mu \mu} (t)$ such that the high-frequency region of the spectra is filtered out. Then, we obtain converged spectra with a reasonably small number of $L$. For the choice of the low-pass filter, we use the Butterworth filter, following the work by Hague \textit{et al.} \cite{MLFT_HaugeRepiskyJCTC23}.   

\section{Implementation}

We have implemented the TD-ASCI algorithm in a developmental version of the \texttt{QChem} quantum chemistry package \cite{epifanovsky2021software}. Using the TD-ASCI time-propagation scheme, we generate the tridiagonal matrix $\mathbf{H}_p$ of dimension ($N_{Krylov} \times N_{krylov}$) and a vector $X$ of dimension ($N_{Krylov} \times 1$), as described in Section \ref{sec:dipole_propagation},
on the sparse time grid defined by the Lanczos algorithm. We store those elements in \texttt{hdf5} file. To perform an FT of the signal, it is necessary to evaluate it on a much finer mesh with a uniform grid spacing. We generate such a signal by exactly interpolating the TD-ASCI signal in between two adaptively located Lanczos nodes, according to the procedure described in Section \ref{sec:dipole_propagation}. Since this procedure involves very small quantities ($\mathbf{H}_p$ and $X$), we perform it with a \texttt{Python} based utility program. The ESPRIT-based FT has also a very modest computing cost, and we implement this in a \texttt{Python} based utility program using \texttt{scipy} and \texttt{numpy} libraries.    

\section{Application} \label{sec:application}

In this section, we demonstrate the key
aspects of the time-propagation algorithm described above through a few representative examples. First, we 
benchmark our method by comparing it 
with the time-propagation of the FCI wave function for a small test system, 
namely for H$_2$O 
in the 6-31G basis set. 
We then study the dynamics of 
several strongly correlated problems 
in molecular systems and evaluate absorption spectra. We will study the convergence of the correlation function and the absorption spectra with respect to the number of core determinants ($N_{cdet}$), 
with the total number of target determinants ($N_{tdet}$) 
chosen to be sufficiently high, 
so that we always reach convergence with respect to this number. 
We analyze the convergence of the FT scheme with respect to the time limit of the propagation, $T_f$.
To obtain the FT of the signal, we exactly interpolate the data obtained from the TD-ASCI algorithm on a much denser uniform grid, with a grid spacing of 0.01 a.u.     

\subsection{Water}

\begin{figure}
    \centering
    \includegraphics[width=0.5\textwidth]{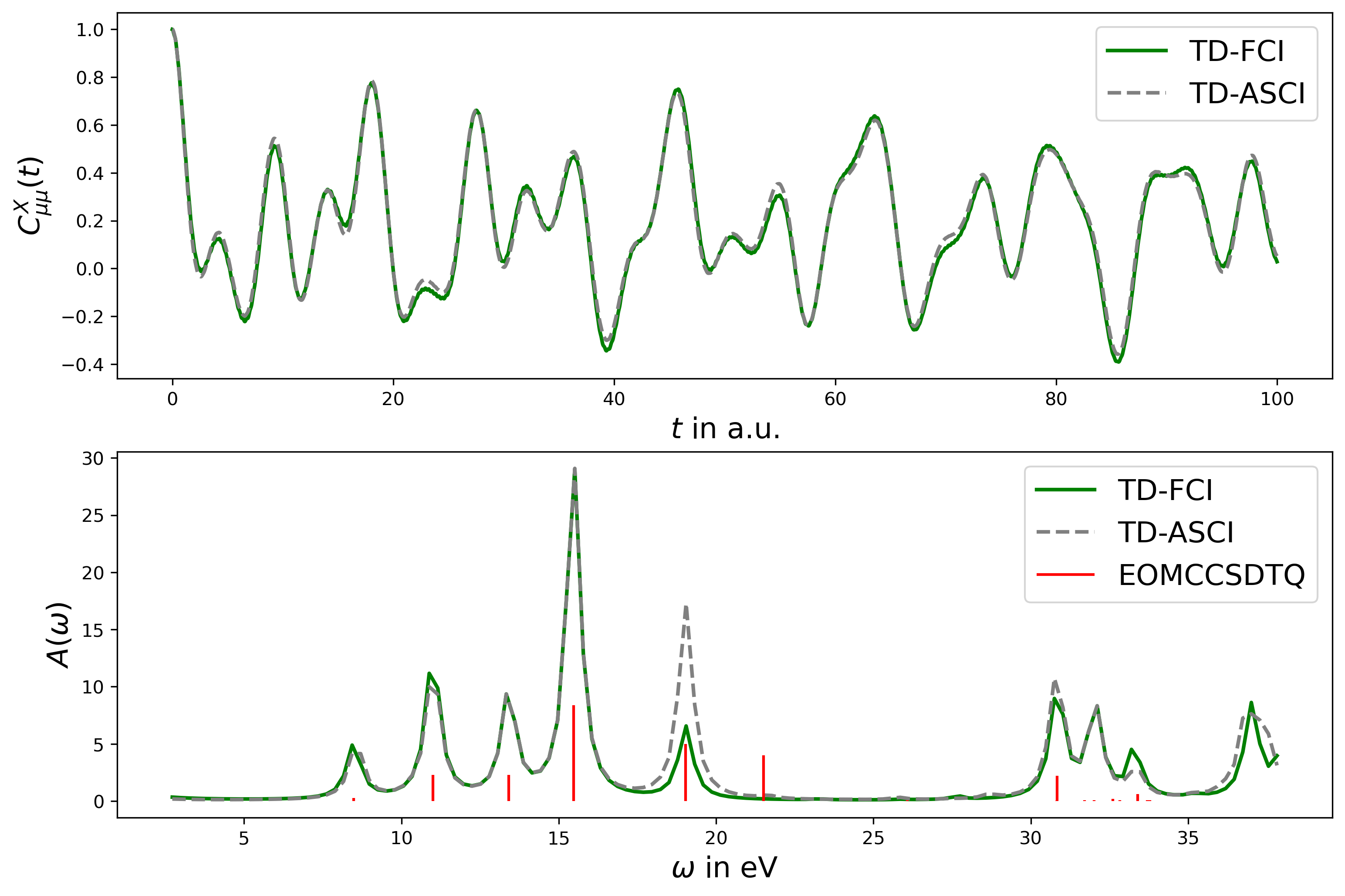}
    \caption{TD-ASCI tests of electronic dynamics for the H$_2$O molecule at its equilibrium geometry in the 6-31G basis set.  (Top panel): the $X$-polarization of the dipole-dipole auto-correlation function is evaluated with the TD-FCI and TD-ASCI methods. $T_f = 100$ a.u. (Bottom panel): The total linear absorption spectrum is calculated by taking the FT of the $X$-, $Y$- and $Z$- polarizations of the dipole-dipole correlation function and summing the result.}
    \label{fig:H2o_benchmark}
\end{figure}

For a first benchmark of the TD-ASCI method, we evaluated $C_{\mu \mu} (t)$ for H$_2$O in the 6-31G basis set with the TD-FCI method and then compared it with the TD-ASCI method developed in this work. We have propagated the dipole vector until 100 a.u. with the time-evolution scheme. For the TD-ASCI time propagation, we chose $N_{krylov}=40$, which amounts to $\sim$ 200 refresh steps for each polarization of the dipole vector. We used 
$N_{cdets} = 4000$ for the ASCI algorithm, which leads to $N_{tdet} \approx 50000$ for the time-propagation. The FT of $C_{\mu \mu} (t)$ is performed with the ESPRIT method described in Sec. \ref{Sec:FT}. For this example, there is no appreciable difference with the Pad{\'e}-based FT (not shown here).
The comparison of the TD-ASCI and TD-FCI is summarized in Fig. \ref{fig:H2o_benchmark}.  
It is apparent that the $C_{\mu \mu} (t)$ signal evaluated with TD-FCI (upper panel) does not differ appreciably from the TD-ASCI signal. This is reflected in the FT of the signal (lower panel) - the peak positions and intensities of the resulting absorption spectra align very well, except the intensity of one peak near the 20 eV region is deviating appreciably. The lower panel of Fig.~\ref{fig:H2o_benchmark} also shows a comparison of the absorption peaks with the EOM-CCSDTQ excitation energies (stick spectra shown as red vertical lines), which we expect to be highly accurate for this case. We find that the peak values agree with the EOM-CC excitation energies to very high precision, with an average deviation of $\sim 10^{-4}$ eV.


\begin{figure}
    \centering
    \includegraphics[width=0.5\textwidth]{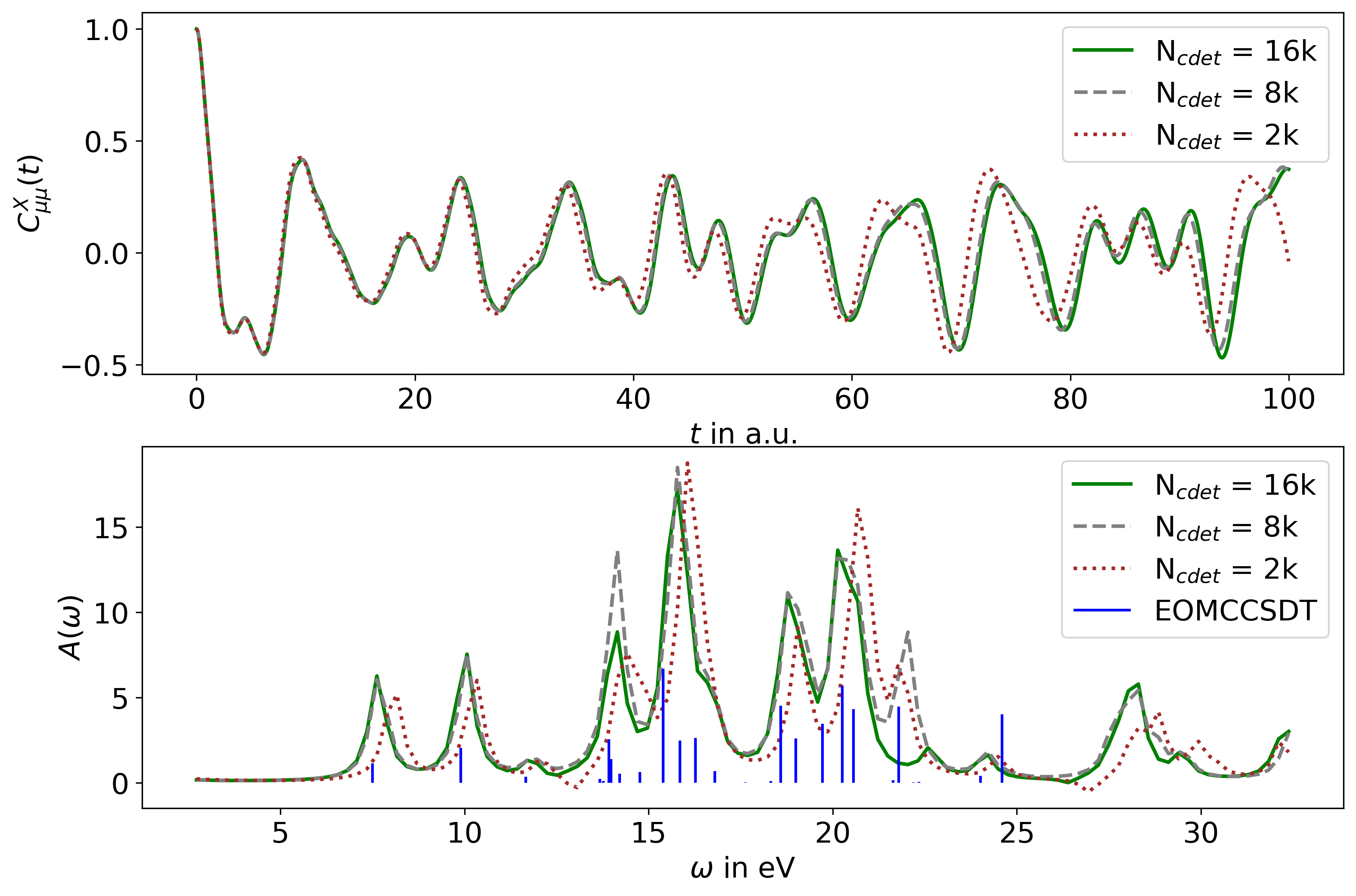}
    \caption{Convergence of TD-ASCI with respect to the $N_{cdet}$ parameter for H$_2$O at its equilibrium geometry, using the aug-cc-pVDZ basis set.
    (Top panel): $X$-polarization of the dipole-dipole auto-correlation function evaluated with the TD-FCI and TD-ASCI methods. $T_f = 100$ a.u. $N_{cdet} = 2000, 8000, 16000$. (Bottom panel): Total absorption spectrum calculated by taking the FT of the $X$-, $Y$- and $Z$- polarizations of the correlation function and summing the result.}
    \label{fig:ncdet_convergence}
\end{figure}

We now evaluate the absorption spectra with a larger basis, namely, the aug-cc-pVDZ basis set, for the same equilibrium geometry of H$_2$O (Fig.~\ref{fig:ncdet_convergence}). For this study, we froze one core orbital of water and chose 40 Lanczos vectors to build the reduced propagator. In this case, we are not able to carry out the FCI calculation, therefore the EOM-CCSDT stick spectrum will be considered as the benchmark reference. We also point out that in this geometry we observe an average deviation of only $\sim$~.04 eV if we truncate the CC hierarchy at the singles-doubles (SD) level instead of SDT. Thus, the underlying wave function in the ground and the excited states may not be strongly correlated.    
In this study, we examine the convergence of the calculated spectra with respect to the various computational parameters of TD-ASCI, namely $N_{cdet}$ and $T_f$. 

To show the convergence with respect to the value of $N_{cdet}$, 
we considered the three values $N_{cdet} = 2000, 8000$ and $16000$. Fig. \ref{fig:ncdet_convergence} summarizes the results. We observe that for both the dipole correlation function $C_{\mu \mu} (t)$ and the absorption spectrum, very good convergence is achieved with $N_{cdet} = 8000$. It is important to observe that although for $N_{cdet} =2000$ the peak positions are not converged, in this case, they nevertheless exhibit a constant shift of $\sim$ +.05 eV relative to the EOM-CC peaks, which  
suggests that a perturbative correction might improve the resulting spectra. We note that for $N_{cdet} =2000$ with perturbative corrections \cite{tubman2018efficientdeterministicperturbationtheory}, the ground state energy is converged within $\sim 10^{-3}$ eV. 

\begin{figure}
    \centering
    \includegraphics[width=0.5\textwidth]{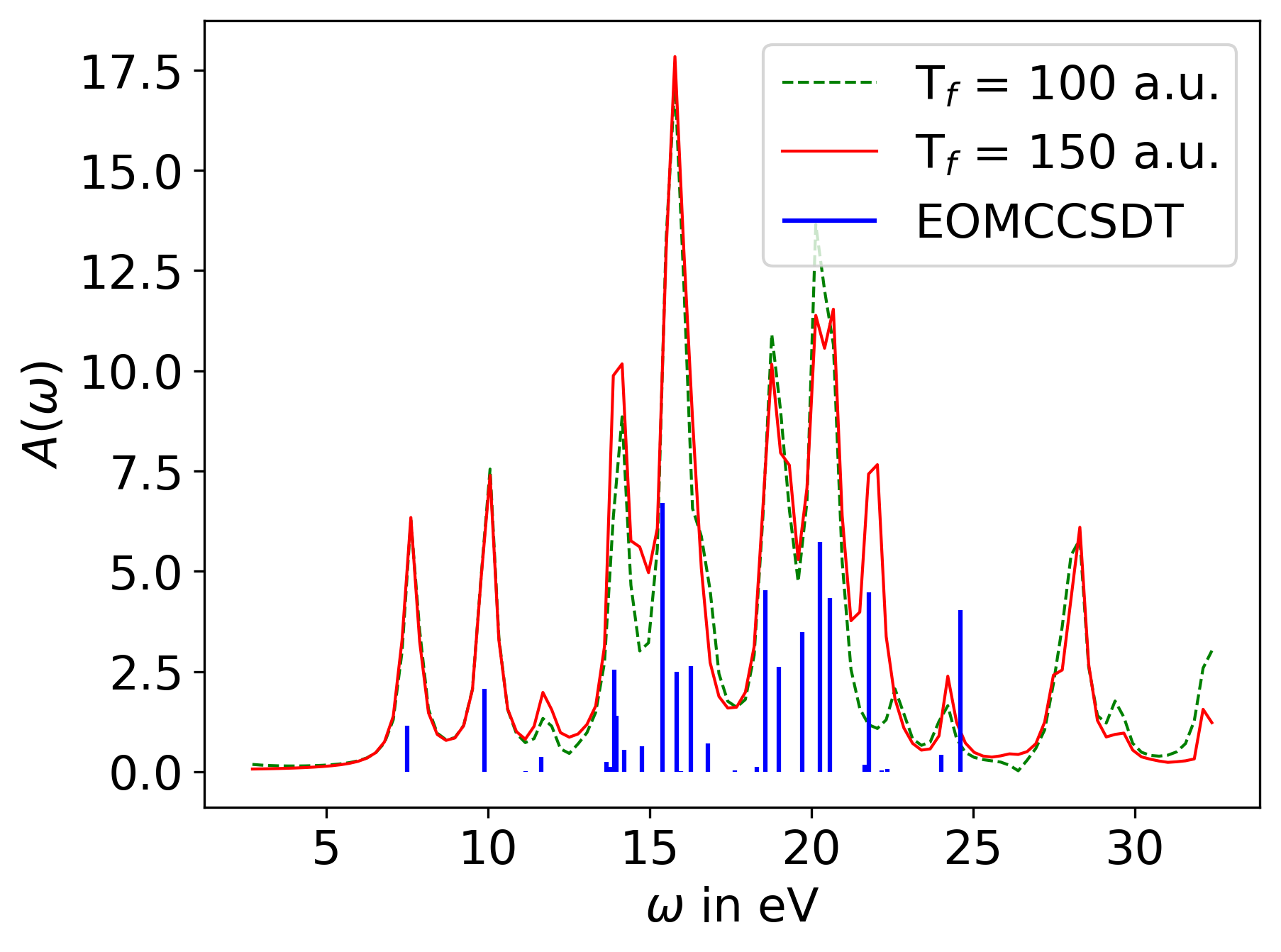}
    \caption{Linear absorption spectrum of H$_2$O molecule at its equilibrium geometry, with the aug-cc-pVDZ basis set, evaluated by TD-ASCI. Results are shown for two different values of the time limit $T_f = 100$ a.u. and $150$ a.u. on the dipole-dipole correlation function $C_{\mu\mu}(t)$ that is input to the FT (Section~\ref{Sec:FT}). $N_{cdet} = 16000$. The spectra are compared with stick spectra obtained from EOM-CCSDT.}
    \label{fig:Tf_conv_SR}
\end{figure}

The influence of imposing a short time limit on the evaluation of the FT, i.e., small values of $T_f$, is analyzed in Fig. \ref{fig:Tf_conv_SR}. For this study, we chose the largest value of $N_{cdet}$, i.e., $N_{cdet} =16000$. 
We have analyzed two values of $T_f$, 100 a.u. and 150 a.u., respectively. The comparison shows that with $T_f = 100$ a.u., we already reach very good convergence in the peak positions and the intensity of the spectral features.        

\begin{figure}
    \centering
    \includegraphics[width=0.5\textwidth]{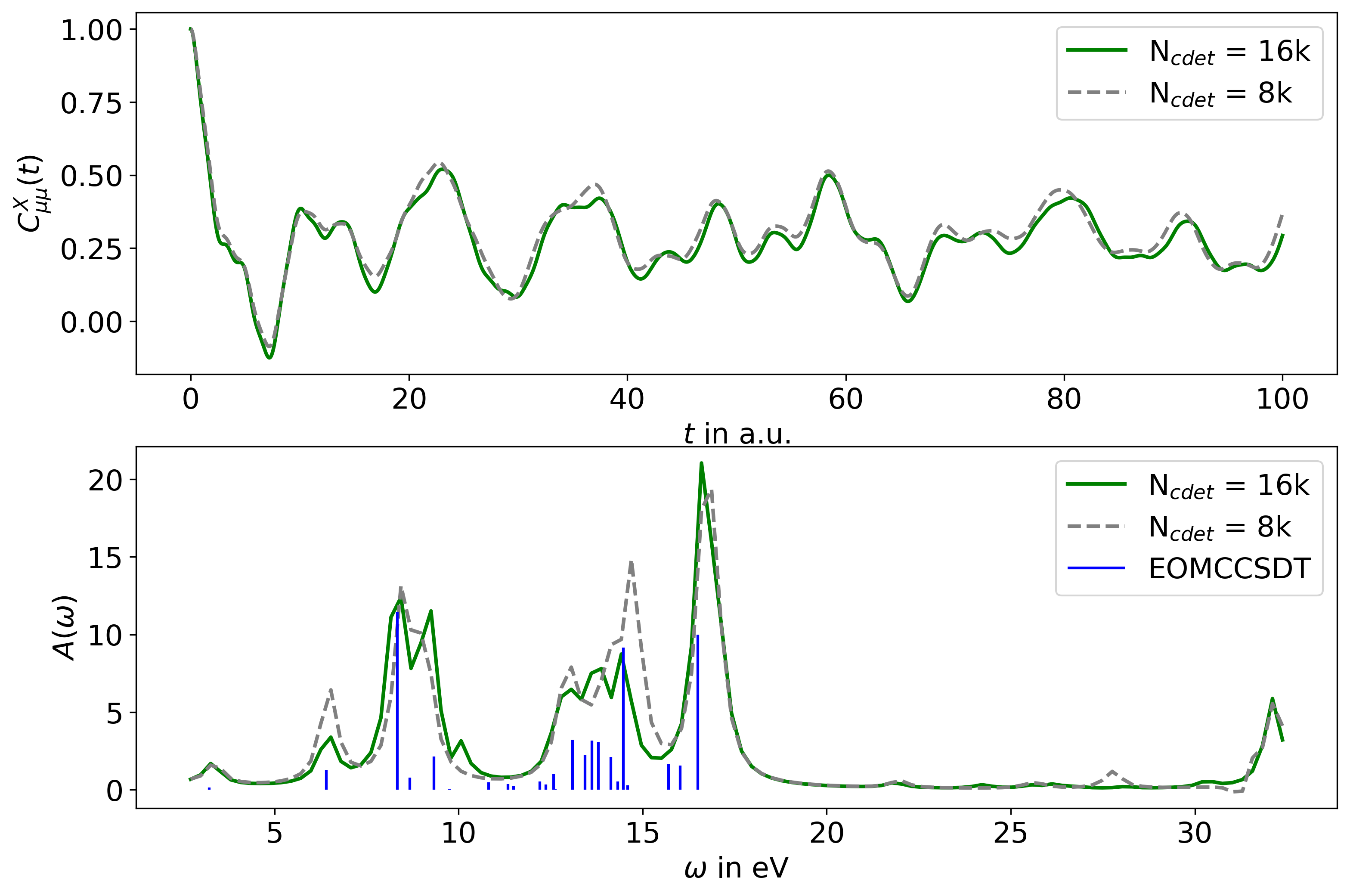}
    \caption{Convergence of TD-ASCI with respect to the $N_{cdet}$ parameter for H$_2$O at a symmetrically stretched geometry with $\angle HOH = 104.52^\circ$, $r_{OH} = 2.6$ a.u., using the aug-cc-pVDZ basis set. (Top panel):  $X$-polarization of the dipole-dipole auto-correlation function evaluated with the TD-ASCI method. $T_f = 100$ a.u. (Bottom panel): The total absorption spectrum is calculated by taking the FT of the $X$-, $Y$- and $Z$- polarizations of the correlation function and summing the result.}
    \label{fig:water_MR_cdets}
\end{figure}

We now compare these spectra with that for a symmetrically stretched geometry of water ($\angle HOH = 104.52^\circ$, $r_{OH} = 2.6$ a.u.). From Tab. \ref{tab:water_MR_peaks} we see a much larger average difference of .2 eV than the equilibrium geometry when we include triples. This suggests that the electronic states are now strongly correlated. For this geometry, we chose $N_{cdet}=8000$ and $16000$. The results are shown in Fig. \ref{fig:water_MR_cdets}. There is now a quite noticeable change in the absorption spectra. For example, in the 5-10 eV spectral region, the calculation with $N_{cdet} = 16000$ shows three distinct peaks, while that with $N_{cdet} = 8000$ shows two peaks.  The difference between the energies obtained with these two parameter values is now not a constant shift so it is not clear whether the calculations are converged at $N_{cdet} = 16000$.

\begin{table}[]
    \centering
    \begin{tabular}{c|c|c|c}
        Molecules & EOM-CCSD & EOM-CCSDT & TD-ASCI  \\
         \hline
         H$_2$O    & 6.63937 & 6.41016 & 6.44387 \\
        (stretched)       & 9.58444 & 9.32519 & 9.29414 \\
                  & 14.61484 & 14.47353 & 14.47544 \\
          \hline
    \end{tabular}
    \caption{Frequency values in eV of three most intense peaks obtained from various methods at a symmetrically stretched geometry of H$_2$O with $\angle HOH = 104.52^\circ$, $r_{OH} = 2.6$ a.u., using the aug-cc-pVDZ basis set.}
    \label{tab:water_MR_peaks}
\end{table}

 We then compare three different lengths of the time propagation: $T_f = 100$ a.u., 250 a.u. and $300$ a.u. Note that the largest value of $T_f$ chosen for this comparison is higher than that for the equilibrium geometry (Figs.~\ref{fig:ncdet_convergence} and \ref{fig:Tf_conv_SR}), since the spectra for the stretched molecule do not converge well even at $T_f = 250$ a.u. It is now evident that propagating for the shorter time $T_f = 100$ a.u. gives rise to a spurious peak near the $\sim$ 10 eV region of the spectra.
 For this shorter time propagation, the intensities of the peaks are also seen to be modified in the 10-15 eV region of the spectrum.

\begin{figure}
    \centering
    \includegraphics[width=0.5\textwidth]{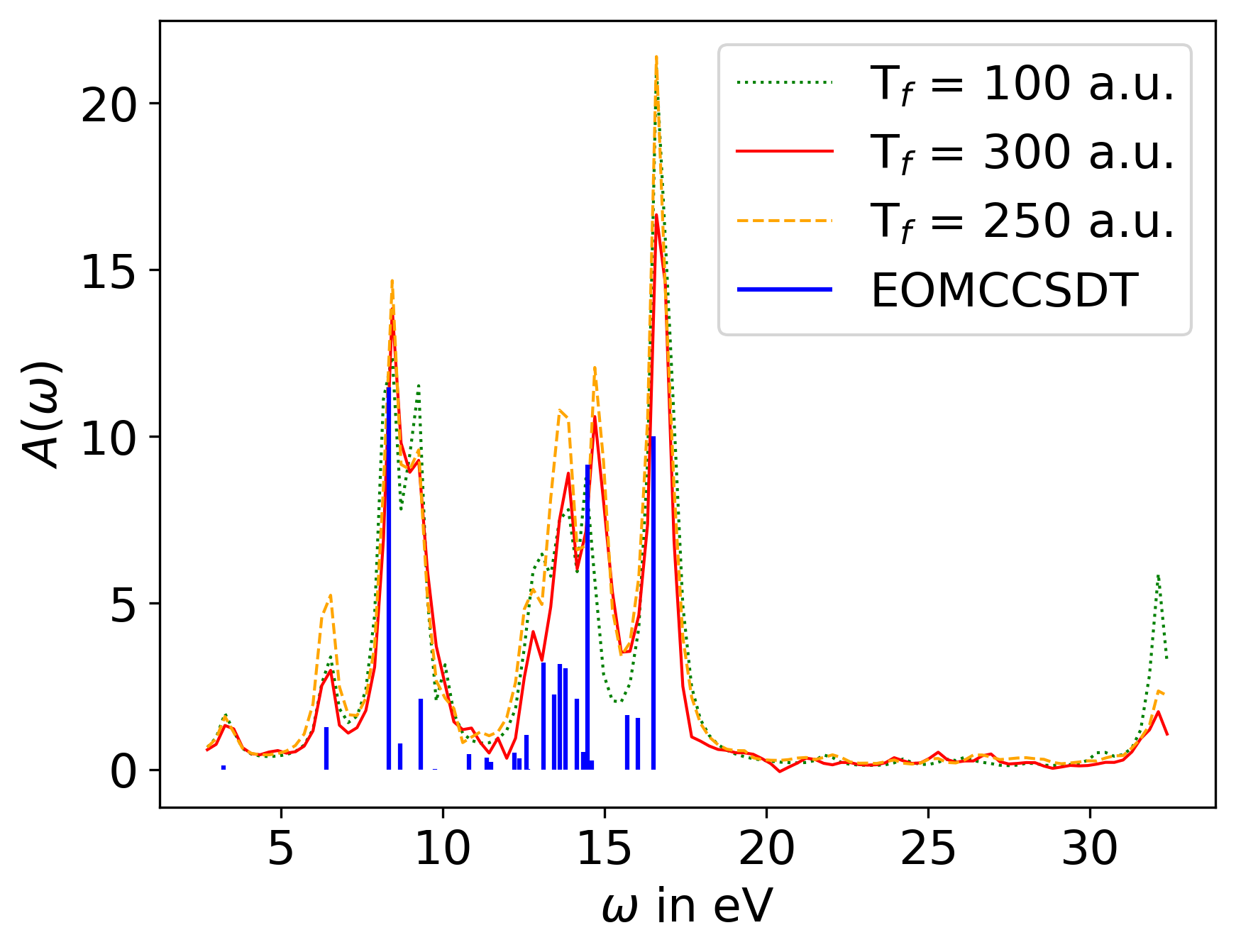}
    \caption{Convergence of the FT spectra with respect to the time limit $T_f$ of the dipole-dipole correlation function calculated with TD-ASCI for H$_2$O  at the symmetrically stretched geometry with $\angle HOH = 104.52^\circ$, $r_{OH} = 2.6$ a.u., using the aug-cc-pVDZ basis set. Results for $T_f = 100$ a.u. and $T_f= 300$ a.u. are shown.}
    \label{fig:tf_conv_MR}
\end{figure}

 The strongly correlated nature of the problem in this geometry is reflected in the fact that the total number of determinants required for the time evolution, i.e., $N_{tdet}$, increases more than twofold (from $N_{tdet} = 521329$ to $N_{tdet}=1105178$ for the $X$-polarization) from its value for the equilibrium geometry.  

\subsection{C$_2$ and CO}
\begin{figure}
    \centering
    \includegraphics[width=0.5\textwidth]{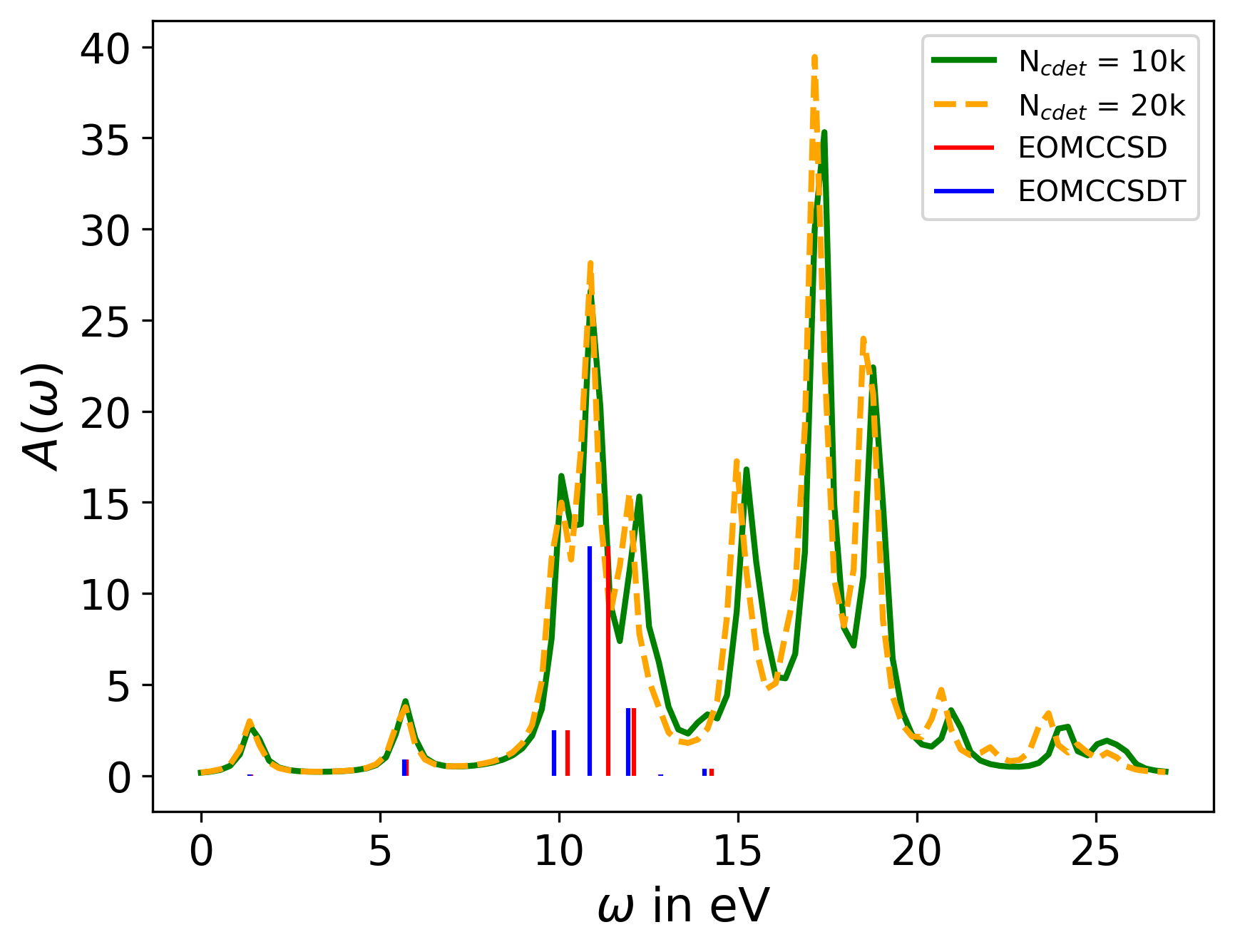}
    \caption{TD-ASCI calculations of the electronic absorption spectrum of the C$_2$ molecule at its equilibrium geometry, using the aug-cc-pVDZ basis set. Results are shown for $N_{cdet}= 10000$ and $N_{cdet}= 20000$. The total absorption spectrum is calculated by taking the FT of the $X$-, $Y$- and $Z$- polarizations of the correlation function and summing the result.}
    \label{fig:c2_cdets}
\end{figure}

In this section, we consider C$_2$ in its equilibrium geometry, which shows strongly correlated behavior for both the ground and excited states. C$_2$ thus serves as a very good prototypical example to study strongly correlated dynamics. For this analysis, we freeze two core-occupied orbitals of C$_2$. We consider two different choices of $N_{cdet}$, namely, $N_{cdet}=10000$ and $20000$. We achieve very good convergence in the resulting absorption spectra using $N_{cdet} = 10000$, as is shown in Fig. \ref{fig:c2_cdets}. By analyzing the EOMCC spectra, we observe a significant discrepancy between results from the SD and singles doubles and triples (SDT) truncations (see Tab. \ref{tab:C2_peaks}). For example, the lowest $\pi-\pi^*$ transition shows a difference of 0.53 eV between the two calculations.   

\begin{table}[]
    \centering
    \begin{tabular}{c|c|c|c}
        Molecules & EOM-CCSD & EOM-CCSDT/CC3 & TD-ASCI  \\
         \hline
         C$_2$    & 10.24819 & 9.86274 & 9.94393 \\
                  & 12.09250 & 11.93477 & 11.92876 \\
                  & 11.37952 & 10.85062 & 10.84929 \\
    \end{tabular}
    \caption{Frequency values in eV of three most intense peaks obtained from various methods at the equilibrium geometry of C$_2$, using the aug-cc-pVDZ basis set.}
    \label{tab:C2_peaks}
\end{table}

\begin{figure}
    \centering
    \includegraphics[width=0.5\textwidth]{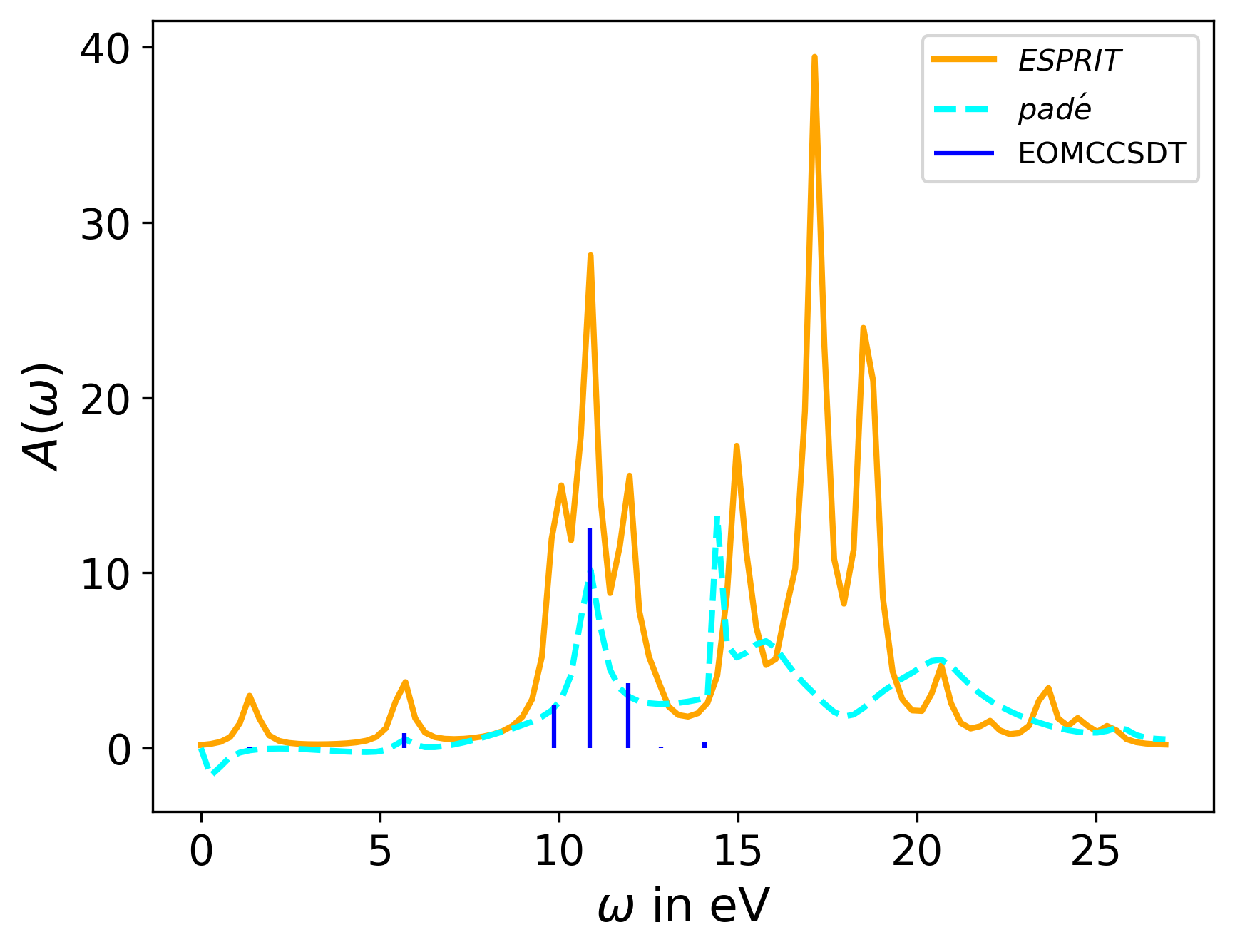}
    \caption{Comparison of TD-ASCI linear absorption spectra of the C$_2$ molecule at its equilibrium geometry, with the aug-cc-pVDZ basis set, obtained by taking the FT of the TD-ASCI dipole-dipole correlation function with i) the Pad{\'e}-based (cyan dashed line) and ii) the ESPRIT-based (yellow solid line). Here $N_{cdet}= 20000$.}
    \label{fig:pade_esprit}
\end{figure}

A comparison of the two FT schemes described in Sec.~\ref{Sec:FT} using $N_{cdet}=20000$ is shown in Fig. \ref{fig:pade_esprit}. We see that the Pad{\'e}-based FT fails to resolve the many peaks lying in the low-frequency region. For example, in the 10-15 eV region of the spectrum, the ESPRIT-based FT reveals three peaks, two of which are of $\pi-\sigma$ character,
and the third is predominantly of $\pi-\pi^*$ character. The Pad{\'e}-based scheme could resolve only the single $\pi-\pi^*$ transition.   

\begin{figure}
    \centering
    \includegraphics[width=0.5\textwidth]{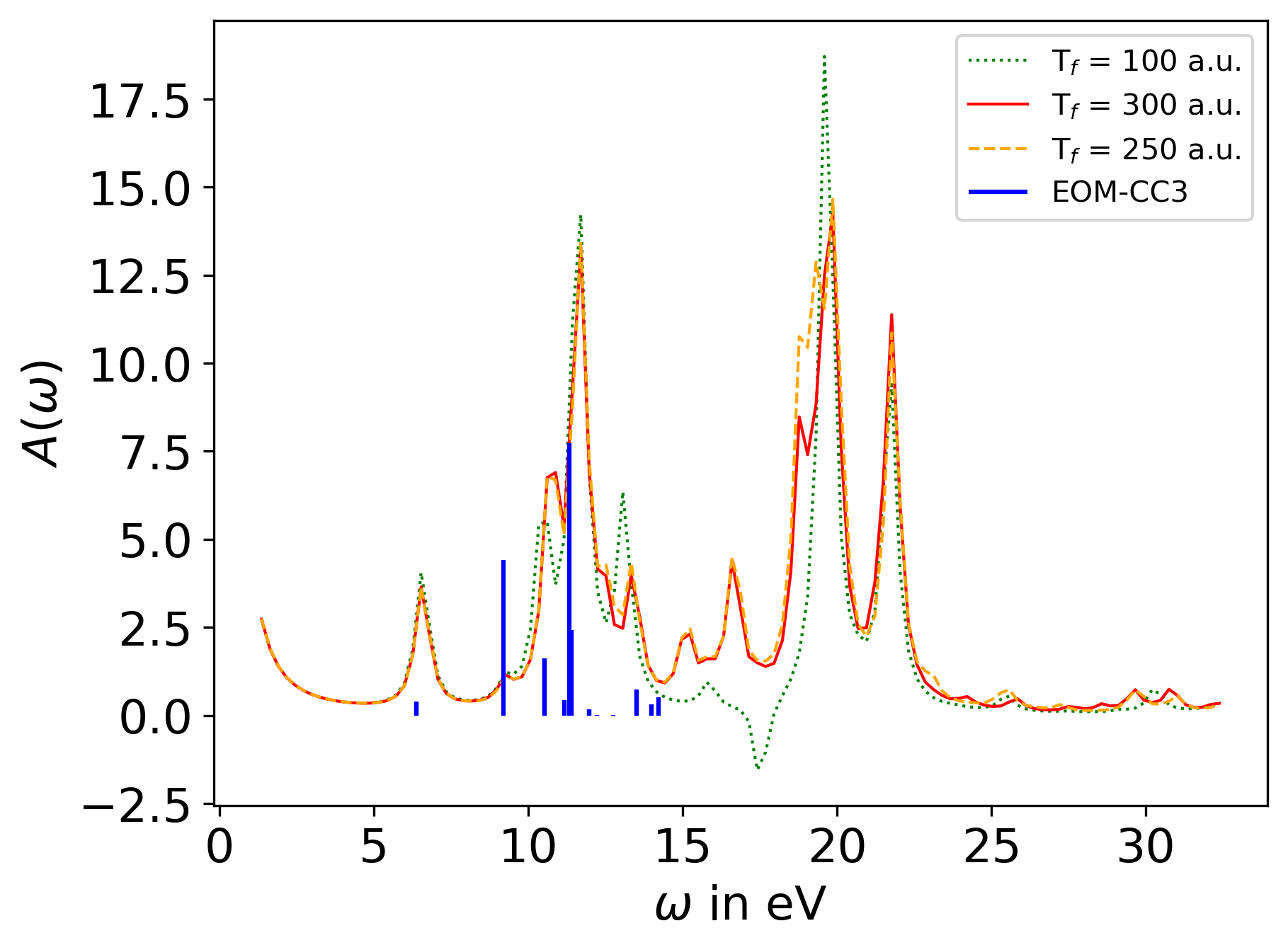}
    \caption{TD-ASCI linear absorption spectrum of the CO molecule at a stretched geometry (1.4 \AA~bond distance), using the aug-cc-pVDZ basis. The spectrum is obtained by taking the FT of the dipole-dipole correlation function. The location of the peaks relative to the EOM-CC stick spectra quantifies the accuracy of the method. Comparison of the spectra shown for three values of the time-limit $T_f$, indicate that the spectrum is converged by $T_f = 250$ a.u.}
    \label{fig:CO_MR}
\end{figure}

Similarly, we have studied the CO molecule in a stretched geometry -- at a bond distance of 1.4 \AA~ where the molecule has multireference character. We first studied how the corresponding spectra converge with the propagation time limit, $T_f$, for which we use 
the following values: 100 a.u., 250 a.u. and 300 a.u. The results are shown in Fig. \ref{fig:CO_MR}, where they are compared with results from EOM-CC3 \cite{EOM_CC3_JCP95} calculations which is an approximation to the EOM-CCSDT method. It is evident that we obtain very well converged spectra at $T_f = 250$ a. u. However, for the shorter overall time-propagation $T_f = 100$, we see unphysical behavior in the 15-20 eV spectral region, with negative values of $A(\omega)$. We also see that in this region the relative intensities of the peaks are modified as we propagate for longer times, indicating a lack of convergence in this regime. 

\begin{figure}
    \centering
    \includegraphics[width=0.5\textwidth]{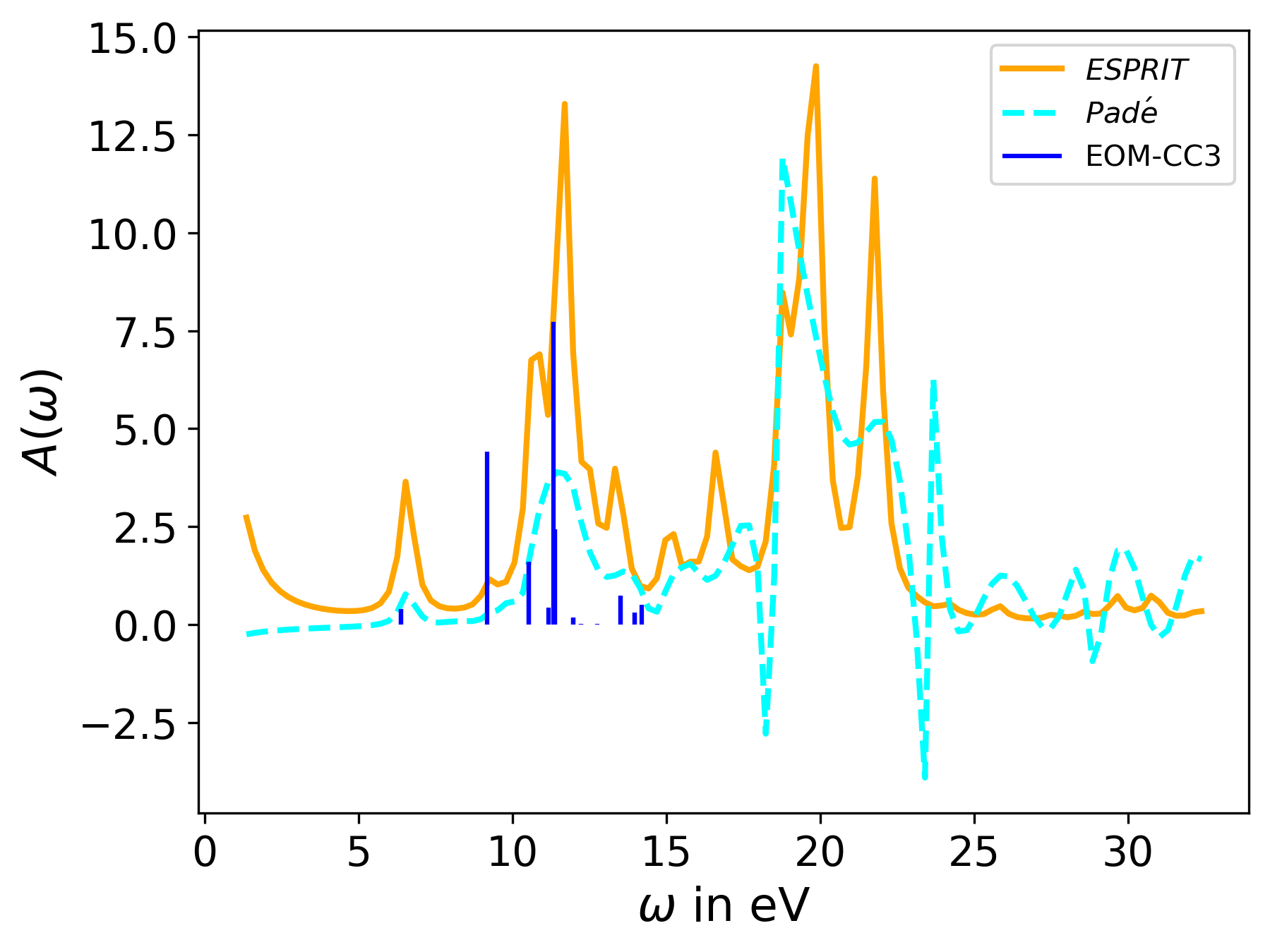}
    \caption{TD-ASCI absorption spectrum for the CO molecule at a stretched geometry (1.4 \AA~bond distance), using the aug-cc-pVDZ basis set. $N_{cdet} = 20000$ here. The performance of the spectra obtained with the Pad{\'e}-based (solid yellow lines) and ESPRIT-based (cyan dashed lines) FT are compared. The solid blue stick spectra represent spectra obtained from EOM-CC3.}
    \label{fig:pade_esprit_CO}
\end{figure}

A comparison for CO on the performance of the two FT schemes 
is shown in Fig. \ref{fig:pade_esprit_CO}.  The FT based on ESPRIT
shows 4 peaks in the 10-15 eV spectral region.
This is corroborated by the number of peaks obtained from the EOM-CC3 spectra, which are reproduced here from Fig.~\ref{fig:CO_MR}. In contrast, the Pad{\'e}-based FT generates only one peak and a shoulder. While it was not possible to characterize peaks beyond the 15 eV spectral region with the EOM-CC3 method, we see that the ESPRIT-based scheme does generate a few distinct peaks in that high-frequency region, while the Pad{\'e} shows unphysical characteristics here.    

\section{Summary and Outlook} \label{sec:conclusion}

We have presented a new wave function-based time propagation algorithm that can take into account the strong correlation behavior of the underlying many-body problem. The algorithm uses an efficient Krylov subspace-based time-integrator, namely the short iterative Lanczos 
method, which produces a norm-preserving and stable long-time propagation scheme by an efficient refreshing technique of the Krylov space. The time propagation algorithm naturally produces an adaptive, sparse time grid, and the cost of the algorithm scales linearly with the size of that grid, $\mathcal{O}(N_t)$, with negligible contribution from interpolation of functions at intermediate times.

The TD-ASCI method was utilized here to evaluate the dipole-dipole autocorrelation function. For this purpose, it was necessary to build a new determinant space for the time propagation by acting with the dipole operator on the ASCI ground state, which yields the single excitation space relative to the ground state. After each variable time intervals $\tau_{max}$, the refreshed Krylov sub-space was constructed from the CI vectors at that time step. The accuracy of the time-correlation function $C_{\mu \mu}(t)$  was estimated by calculating the absorption spectrum from its FT and comparing this with the spectrum from EOM-CCSDT or EOM-CC3 calculations. A new FT scheme based on the signal processing algorithm ESPRIT enabled us to extract very accurate spectra from a short-time signal, thereby significantly reducing the overall computational cost of the time propagation. We have shown empirically that the new FT scheme performs significantly better in resolving dense spectra than the well-known Pad{\'e}-based FT.
Future work will seek to develop a rigorous error analysis with respect to the time limit for convergence. 

We have demonstrated the use of the TD-ASCI method here with an evaluation of the absorption spectra of H$_2$O and CO in stretched geometries, and of C$_2$ in its equilibrium geometry, using these small systems as prototypical examples of strongly correlated system. The accuracy of the method was demonstrated by showing the proximity of the TD-ASCI spectral peaks to theoretical estimates of the corresponding peaks obtained from accurate linear response theories, specifically from EOM-CCSDT or EOM-CC3. We have observed that the present time-propagation scheme requires a large number of $N_{cdet}$, i.e., the number of ground state core determinants from which the new determinant space suitable for time propagation is built. This can be attributed to the lack of perturbative corrections to the calculated spectra. Future work will address schemes to incorporate such corrections and thereby allow a reduction in $N_{cdet}$. Another possible way to reduce $N_{cdet}$ is to target only a finite number of low-energy absorption peaks that lie within a specific energy window. In this situation, we expect that it will be possible to drastically truncate the search space. 

It is also straightforward to evaluate other correlation functions within the current framework. One such example is the quadrupole-quadrupole correlation function, which will allow us to explore spectra beyond dipole-allowed transitions. We will also investigate the applicability of the current method to a Hamiltonian with an explicitly time-dependent driving term, to go beyond the linear response regime, e.g., to consider quench dynamics.

A major advantage of the TD-ASCI method lies in the fact that for a given strongly correlated problem, we can obtain a very accurate description of the dynamics. While evaluating the full system dynamics of a realistic electronic system with very large numbers of electrons using the TD-ASCI method would be numerically intractable, the method appears ideally situated to study the real-time dynamics of electronic systems within the framework of reduced dynamics and open quantum systems (OQS) dynamical methods.
One promising application is in the direction of reduced dynamics, with the fermionic impurity problem. Such an approach will allow the study of quantum dots \cite{quantumdot_PRL17, goldhaber1998kondo}, nanowires \cite{nanowire_PhysRevLett98} and molecular junctions \cite{aradhya2013single}. Another promising application is in the direction of OQS dynamics, namely the non-Markovian dynamics of a reduced electronic system. Typical OQS formulations divide many-body dynamical problems into a strongly interacting, non-Markovian system that is treated explicitly and a bath that is weakly coupled to the system and that can be either Markovian or non-Markovian. TD-ASCI is well-suited for describing dynamics of the strongly interacting, non-Markovian system, within the general frameworks of reduced open quantum system dynamics such as the hierarchy of pure states (HOPS) \cite{HOPS_PhysRevLett2014}, the hierarchical equations of motion (HEOM) \cite{tanimura1989time, tanimura2020numerically}, as well as the more restricted Lindblad bath dynamics \cite{lindblad1976generators}. Several recently proposed methods for OQS have combined either matrix product states (MPS) or complete active space (CAS) CI with HOPS \cite{HEOMDMRG_PRL22} or with the Lindblad \cite{LindbladEmbedPRL18, TDCI_impurity_solver_PRB23} dynamics method. The TD-ASCI method offers an alternative approach that provides efficient handling of sparsity, with no limitations in terms of the dimensionality of the interaction. We expect these directions will generate useful extensions of the basic TD-ASCI method that we have presented here.

\section{Acknowledgements}

This material is based upon work supported by the U.S. Department of Energy, Office of Science, Office of Advanced Scientific Computing Research and Office of Basic Energy Sciences, Scientific Discovery through Advanced Computing (SciDAC) program under Award Number DE‐SC0022198 (A.S., K.B.W.), and Award Number DE-SC0022364 (M.H-G.), and is supported by the Simons Targeted Grants in Mathematics and Physical Sciences on Moir\'e Materials Magic (Z.H.). 
This research used resources of the National Energy Research Scientific Computing Center, a DOE Office of Science User Facility supported by the Office of Science of the U.S. Department of Energy under Contract No. DE-AC02-05CH11231 using NERSC award BES-ERCAP0029462.

\pagebreak
\appendix 

\section{Lanczos Tridiagonalization:} \label{sec:lanczos_detail}

\begin{description}
    \item[Step 1] Initialize the Lanczos procedure: $\beta_0  = 0$; $\mathbf{q}^0  = 0$; $\mathbf{q}^1 = \mathbf{C}(t)/\|\mathbf{C}(t)\|$
    \item [step 2] for $j = 1, 2, ... ,N_{Krylov}$, 
  \begin{align*}
    \mathbf{v} = {} & \mathbf{H} \mathbf{q}^j \\
    \alpha_j = {}& (\mathbf{q}^j)^\dagger\mathbf{v}  \\
    \mathbf{v} ={}& \mathbf{v} - \beta_{j-1} \mathbf{q}^{j-1} - \alpha_j \mathbf{q}^j \\
    \beta_j = {}& \lvert\lvert \mathbf{v} \rvert\rvert \\
    \mathbf{q}^{j+1} = {}& \mathbf{v}/\beta_j 
    \end{align*}                    
\end{description}











\bibliography{misc}

\end{document}